\documentclass[12pt,preprint]{aastex}

\slugcomment{Accepted by ApJ}

\shorttitle{The environment of FS Tau observed with HST WFPC2}
\shortauthors{Woitas et al.}

\begin{document}

%% LaTeX will automatically break titles if they run longer than
%% one line. However, you may use \\ to force a line break if
%% you desire.

\title{The environment of FS Tau observed with HST WFPC2 \\ in narrow-band
       emission line filters}

%% Use \author, \affil, and the \and command to format
%% author and affiliation information.
%% Note that \email has replaced the old \authoremail command
%% from AASTeX v4.0. You can use \email to mark an email address
%% anywhere in the paper, not just in the front matter.
%% As in the title, you can use \\ to force line breaks.

\author{Jens Woitas, Jochen Eisl\"offel}
\affil{Th\"uringer Landessternwarte Tautenburg, Sternwarte 5,
       D-07778 Tautenburg, Germany}

\email{woitas@tls-tautenburg.de}
\email{jochen@tls-tautenburg.de}

\author{Reinhard Mundt}
\affil{Max-Planck-Institut f\"ur Astronomie, K\"onigstuhl 17,
       D-69117 Heidelberg, Germany}
    
\email{mundt@mpia-hd.mpg.de}

\and

\author{Thomas P. Ray}
\affil{School of Cosmic Physics, Dublin Institute for Advanced Studies, \\
       5 Merrion Square, Dublin 2, Ireland}

\email{tr@cp.dias.ie}

%% Notice that each of these authors has alternate affiliations, which
%% are identified by the \altaffilmark after each name.  Specify alternate
%% affiliation information with \altaffiltext, with one command per each
%% affiliation.

%% Mark off your abstract in the ``abstract'' environment. In the manuscript
%% style, abstract will output a Received/Accepted line after the
%% title and affiliation information. No date will appear since the author
%% does not have this information. The dates will be filled in by the
%% editorial office after submission.

\begin{abstract}
We present the results of HST WFPC2 observations of FS~Tau and its
environment obtained in the narrow-band emission line filters H$\alpha$ and
[S\,II] $\lambda\lambda\,6716, 6731\,\mathrm{{\AA}}$. Based on these data
the morphology of line emission within this region can be studied on a size
scale of $0\farcs1$ for the first time.\\
Despite the fact that FS~Tau~A has strong forbidden emission lines,
there is no evidence for extended emission at these wavelengths beyond
$0\farcs5$ from the components of this close T Tauri binary system.
In the FS~Tau~B outflow interesting morphological properties can be
studied at high spatial resolution.
In this jet we find a fine structure where circular or elliptical jet knots
are correlated with mimima of the jet width. The overall width of this jet
decreases with distance from the source. The FS~Tau~B jet is thus a rare
example of a jet which may be recollimated far away from its source.
The jet is much more prominent in H$\alpha$ than in [S\,II], while the
counterjet shows the opposite behaviour. The line ratio H$\alpha$/[S\,II]
increases with lateral distance from the jet axis, which is indicative
of entrainment of ambient material.
\end{abstract}

%% Keywords should appear after the \end{abstract} command. The uncommented
%% example has been keyed in ApJ style. See the instructions to authors
%% for the journal to which you are submitting your paper to determine
%% what keyword punctuation is appropriate.

%% From the front matter, we move on to the body of the paper.
%% In the first two sections, notice the use of the natbib \citep
%% and \citet commands to identify citations.  The citations are
%% tied to the reference list via symbolic KEYs. The KEY corresponds
%% to the KEY in the \bibitem in the reference list below. We have
%% chosen the first three characters of the first author's name plus
%% the last two numeral of the year of publication as our KEY for
%% each reference.

\section{Introduction}
The T Tauri star FS~Tau~A (other designations HBC 383,
Haro 6-5 A) is located in the Taurus-Auriga star forming region (SFR)
at a distance of $\approx 140\,\mathrm{pc}$
(e.\,g. Wichmann et al.\,1998). The object and its environment show a
variety of phenomena typical for SFRs. FS~Tau~A is a close binary
with a projected separation of $\approx 0\farcs25$
(Simon et al.\,1992). The young stellar object FS~Tau~B
(other designations HBC 381, Haro 6-5 B) is located
$\approx 20''$ west of FS~Tau~A. It is associated with
a Herbig-Haro jet discovered by Mundt et al.\,(1984).
This jet (HH 157) has a projected length of $\approx 6'$ and a position angle
of $56^{\circ}$ (Eisl\"offel \& Mundt 1998, hereafter EM98). Since the radial
velocities are only $\approx 60\,\mathrm{km\,s^{-1}}$, the jet is probably
orientated quite close to the plane of the sky (Mundt, Brugel, \&
B\"uhrke 1987). There is also a redshifted counterjet that is much less
prominent. FS~Tau~A and B are surrounded by a complex structure of reflection
nebulosities.\\

Krist et al.\,(1998, hereafter K98) have used the Wide Field Planetary Camera
(WFPC2) of the Hubble Space Telescope (HST) to obtain images of the FS Tau
field in the broad-band filters F675W (WFPC2 R band) and F814W (WFPC2 I band).
Based on these data they have examined the spatial structure of the 
reflection nebulosities around FS~Tau A and B. In this paper we present data
obtained with WFPC2 in the narrow-band emission line filters H$\alpha$ and
[S\,II] $\lambda\lambda\,6716, 6731\,\mathrm{{\AA}}$. 
We will focus on properties of the FS~Tau~B outflow
that has been observed for the first time with a spatial resolution of
$\approx 0\farcs1$ in these emission lines. Details of observations and
data reduction will be given in Sect.\,\ref{obsdata}. The results will
be presented in Sect.\,\ref{results} and discussed in Sect.\,\ref{discussion}.

\section{Observations and Data Reduction}
\label{obsdata}
FS Tau and its surroundings were observed with HST WFPC2 on
1997 August 26. In the narrow-band filter F656N (H$\alpha$) one 1400\,s
and two 1300\,s exposures were obtained. Another four images with
integration times of 1300\,s each were taken in the F673N filter
that contains the [S\,II] emission lines at
$\lambda\lambda = 6716, 6731\,\mathrm{{\AA}}$. Exposures in the broad-band
filters F569W and F791W ($3\times 100\,\mathrm{s}$ each) were used for
continuum subtraction. FS~Tau~A, its surrounding reflection nebula and
FS~Tau~B were placed onto the PC chip that provides the
best spatial resolution (pixel scale 0\farcs046/pixel).\\

The basic data reduction used the standard HST data pipeline described
by Holtzman et al.\,(1995a, 1995b), which was
re-run with updated calibration files. The {\tt crrej} task in STSDAS
has been used to remove cosmic rays by combining all images taken
in the same filter. After correcting for the bad pixels marked in the
data quality file a large number of hot or dark pixels remained.
Therefore, additional (automatic and manual) bad pixel corrections
have been applied. Finally, the images taken with the PC and the three WF
chips were mosaiced into one frame. For this purpose the PC images were
re-scaled to fit the pixel scale of the three WF chips
($0\farcs1\,\mathrm{pixel}$). The resultant images in H$\alpha$ and [S\,II]
are shown in Fig.\,\ref{bilder}.

\section{Results}
\label{results}

\subsection{FS~Tau~A}
\label{fstaua}
One goal of our work was to study the spatial structure of 
line emission close to the components of the FS~Tau~A binary.
It was already mentioned by Cohen \& Kuhi (1979) that this system shows
equivalent widths EW(H$\alpha$) = 57.3\,{\AA} and EW([O\,I]) = 21.8\,{\AA}.
Hirth, Mundt, \& Solf (1997) have also detected strong emission in
[O\,II], [N\,II] and [S\,II] associated with this binary system.
White \& Ghez (2001) have resolved the FS~Tau~A binary in H$\alpha$
and found that both components show equivalent widths
EW(H$\alpha$) $>$ 10\,{\AA} which classifies them as classical T Tauri stars
that are supposed to show signs of active disk accretion and bipolar outflows
according to standard models of young stars and their environment
(e.\,g. Hartmann 1998).\\

For a further investigation of this line emission we have used the images
taken in the broad-band filters F569W and F791W for continuum subtraction. 
This subtraction is not straightforward because there are strong colour
gradients in the reflection nebulae. As the probable best solution we scaled
the continuum images in such a way that the large reflection nebula
(labeled with R1 in Fig.\,\ref{bilder})
northeast of FS~Tau~B disappears after subtraction. This removes the
nebulosity around FS~Tau~A down to a projected distance of 
$\approx 0\farcs5$ from the binary components. Neither outside nor inside
this radius do we find any structure indicative of bipolar lobes or
small-scale jets.  

\subsection{The outflow of FS~Tau~B}
\label{fstaub}
In this section we will discuss morphological properties of the
FS~Tau~B jet. Compared to previous studies of this outflow based
on images taken from the ground (e.\,g. Mundt, Ray, \& Raga 1991, hereafter
MRR91, EM98) the spatial resolution is increased by an order of magnitude
using HST. Using the {\it Tiny Tim} software (Krist 1995) we have calculated
model PSFs at the locations of the jet on the WFPC2 chips for the filters
used.
They have FWHMs of $\approx 0.1''$, which is comparable to the adopted pixel
scale. Therefore, the images have not been deconvolved with any PSF.
Since the jet is neither visible in the F569W filter, nor in the
F791W, no continuum subtraction has been applied for this analysis either.\\

In the upper panels of Figs.\,\ref{hamorph} and \ref{siimorph}
contour plots of the outflow region in H$\alpha$ and [S\,II]
are given. The jet source (labeled  as FS~Tau~B in Fig.\,\ref{bilder})
is not visible at optical wavelengths. Following EM98, we assume the
source position to be located $0\farcs8$ southwest of the nearby bright
emission knot. This knot is a strong continuum source,
but remains visible in H$\alpha$ and [S\,II] also after the continuum
subtraction described in Sect.\,\ref{fstaua}. K98
came to the conclusion that this object, and the emission knot
next to it in the southwestern direction, are the upper and lower surfaces
of an edge-on seen disk around FS~Tau~B that is centrally illuminated and
seen in scattered light from the star. In our continuum images only the
bright northeastern knot can be seen that has the same shape as on the
images presented by K98. Concerning our data, the
southwestern knot is only visible in H$\alpha$ and [S\,II]. In the K98
data this object is prominent only in the F675W filter that contains
the [O\,I], H$\alpha$, [N\,II] and [S\,II] lines. On the other hand,
in the F814W filter that only contains weak emission lines like
[Fe\,II]\,$\lambda\lambda 7688, 8617\,\mathrm{\AA}$ it is not significantly above the
background level despite the fact that the integration time was more than
twice as large as for our F791W observation (700\,s compared to 300\,s).
This has been checked by downloading the WFPC2 data used by K98 from the
archive. So this object is probably rather a jet knot than part of a
circumstellar disk.\\

The pear-shaped object close to
FS~Tau~B in the northeastern direction is a reflection nebula (called R1 by
EM98) and is therefore not included in the analysis of
the jet. In the northeastern direction the jet is invisible over a distance
from the source of $d \approx 20''$ in H$\alpha$ and $d \approx 35''$
in [S\,II]. In H$\alpha$ it reappears showing a structure resembling 
a bowshock.
The intensity maximum of this structure (called knot B by MRR91 and EM98)
is at a distance of $d = 22\farcs7$ from the source. According to EM98
this maximum appeared at $d = 18\farcs8$ in November 1990. This shift
of $3\farcs9$ within 6.75 years is roughly in line with the proper motion
of $0\farcs504\,\mathrm{yr^{-1}}$ derived by EM98 for this jet knot.
Following the jet in its downstream direction we can resolve the structure
called A by MRR91 and EM98 (see also Fig.\,\ref{bilder}) into six intensity
maxima or jet knots that have a circular or elliptical shape.
Note that a large portion of
the more distant part of the outflow is not inside the field of view of
WFPC2: There are additional jet knots up to a distance of $\approx 6'$
from the source (e.\,g. EM98).\\

In [S\,II], the northeastern part of the jet is much less prominent
than in H$\alpha$. Here, the [S\,II] flux is roughly an order of
magnitude smaller than the flux in H$\alpha$.  The counterjet shows a
completely different behaviour.  It is brighter in [S\,II] than in H$\alpha$,
and there is even an additional emission knot to the southwest visible
in  [S\,II] which is not seen in H$\alpha$. \\

To perform a quantitative analysis of the outflow morphology we used
a method similar to the one presented by Raga, Mundt, \& Ray (1991).
We have fitted Gaussian curves to the intensity distribution perpendicular
to the jet. In the jet direction the images were usually binned over four
pixels. From the Gaussians we obtained the position of the jet centre, the jet
width and the intensity at the jet centre. The latter is converted from
ADUs to flux using the conversion factors provided by STScI. 
The same fits were subsequently done for single image columns (without
binning). From averaging over the results we have estimated errors for
the jet width and the position of the jet centre.\\

The results are shown in Figs.\,\ref{hamorph} and \ref{siimorph} (the three
lower panels). Between $d=20''$ and $d=33''$ a significant shift of the jet
centre in the direction perpendicular to the flow axis can be seen in
H$\alpha$. The position angle changes by $\approx 10^{\circ}$ over this range.
Another slight change in the outflow direction at $d\approx 35''$  can be
seen in both filters (Figs.\,\ref{hamorph} and \ref{siimorph}, second panel).
Between $d=26''$ and $d=40''$ variations of the jet width as function of
distance to the source can be seen. They form a pattern where minima of
the jet width correspond to jet knots or flux maxima. If only the maxima
or minima of this fine structure are considered, the overall jet width
decreases in this region. The same behaviour can also be seen in [S\,II]
although it is not significant with respect to the error bars here. In that
filter the jet width is about a factor two smaller than in H$\alpha$. 

\section{Discussion}
\label{discussion}

We have found that the jet width decreases with distance from the source
over a large portion of the visible flow as has already been reported by
MRR91. A comparison with other morphological studies of Herbig-Haro jets
shows that this behaviour is rather unusual: MRR91 have examined nine
outflows and found that the jet width usually increases with
distance from the source. The only exceptions from this general behaviour were
HH 24 G and FS~Tau~B in which the jet width decreases. For the FS~Tau~B jet
this behaviour is now confirmed by our much higher spatial resolution
HST images. MRR91 put forward an idea to explain the unusual appearance of
the FS~Tau~B outflow: They suggest that after an initial phase of strong
expansion the jet gets recollimated.\\

Recollimation of disk winds by the hoop stresses produced by toroidal
magnetic fields has been proposed as
a mechanism for focusing YSO jets (e.\,g. Ouyed \& Pudritz 1997).
This process happens however at distances of $\approx 10$\,AU from the
star, whereas for the FS~Tau~B jet the distance between the source and
the first optical jet knot is $\approx 3000$\,AU. Therefore, in this case the
appearance of the outflow is probably the result of an interaction between
the jet
and its external medium: First, the jet overexpands because it moves into a
region of very low pressure. Afterwards, it is recollimated by its ambient
medium. This interaction produces shocks that make the jet visible again at a
projected distance of $\approx 20''$ from the source. This idea could
explain the observed H$\alpha$ bow where the flow becomes visible again.\\

Numerical simulations of such a process have been given by
de Gouveia dal Pino, Birkinshaw, \& Benz (1996) and de Gouveia dal Pino
\& Birkinshaw (1996). They have shown that a morphological structure
very similar to that of the FS~Tau~B jet can appear if the density of
the ambient medium increases with distance from the source. On the other
hand, their simulations produce only very weak shocks if the jet is moving
through a region with decreasing density. This could explain the large
gap between the source and the first optically visible part of the jet.
If the appearance of the FS~Tau~B jet is interpreted with respect to
the mentioned simulations it will primarily reflect density gradients
in the ambient medium that are not spherically symmetric around
the source. This idea is in line with our finding that the southwestern
counterjet is totally different from the northeastern outflow as well
in morphology as in its excitation. It also supports the conclusion of
K98  that the FS~Tau~B jet passes through a ``dark nebula''
northeast of the source.\\

Another possible mechanism for recollimating the FS Tau B jet is provided
by the recent hydrodynamic (Delamarter, Frank, \& Hartmann 2000) and
magneto-hydrodynamic (Gardiner, Frank, \& Hartmann 2001) simulations of
YSO winds propagating into infalling environments. In such simulations not
only does the ambient medium help to recollimate any expanding wind through
shock-refocusing but focusing is enhanced through the ram pressure of the
surrounding medium.\\ 

\begin{figure}
\plottwo{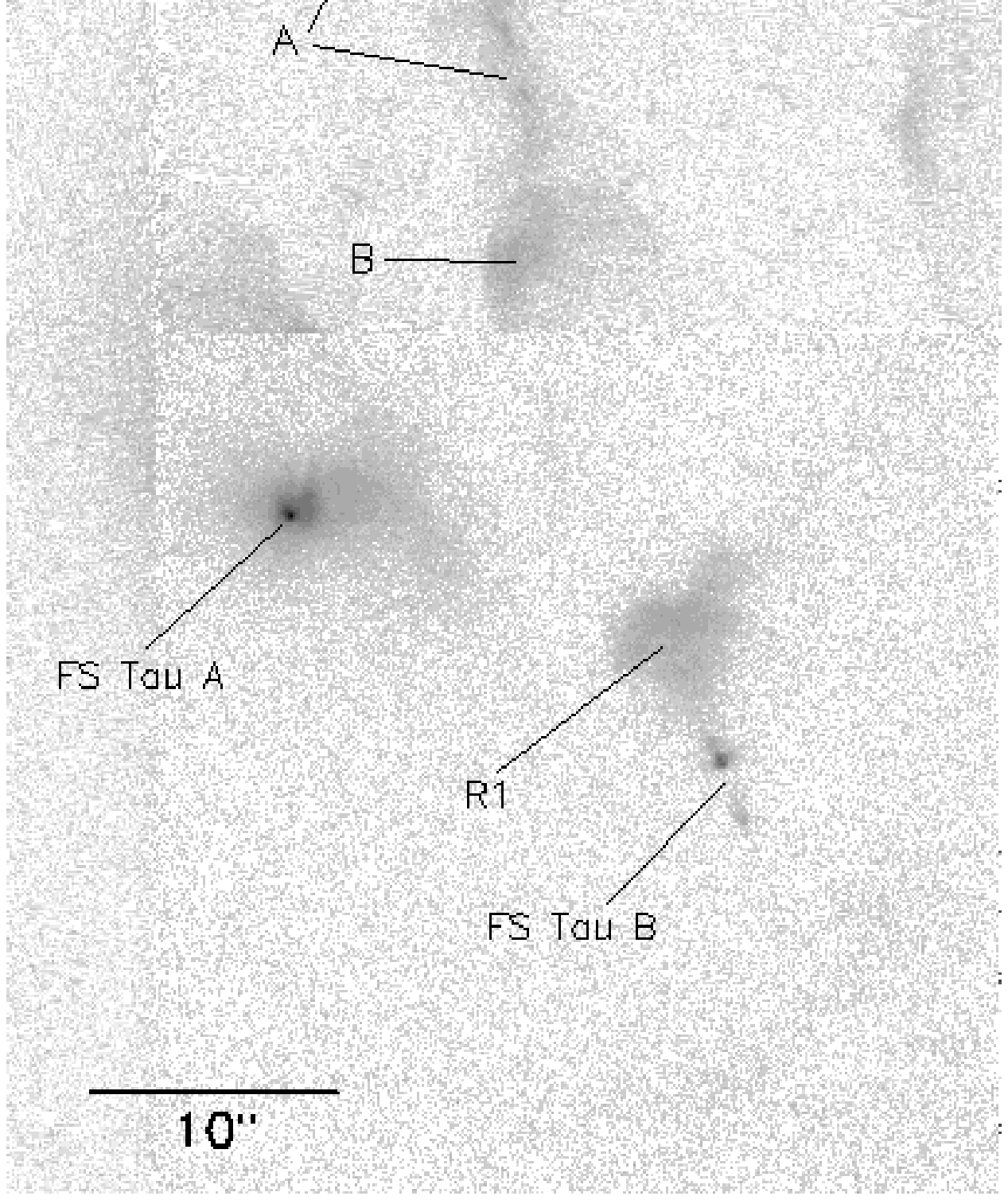}{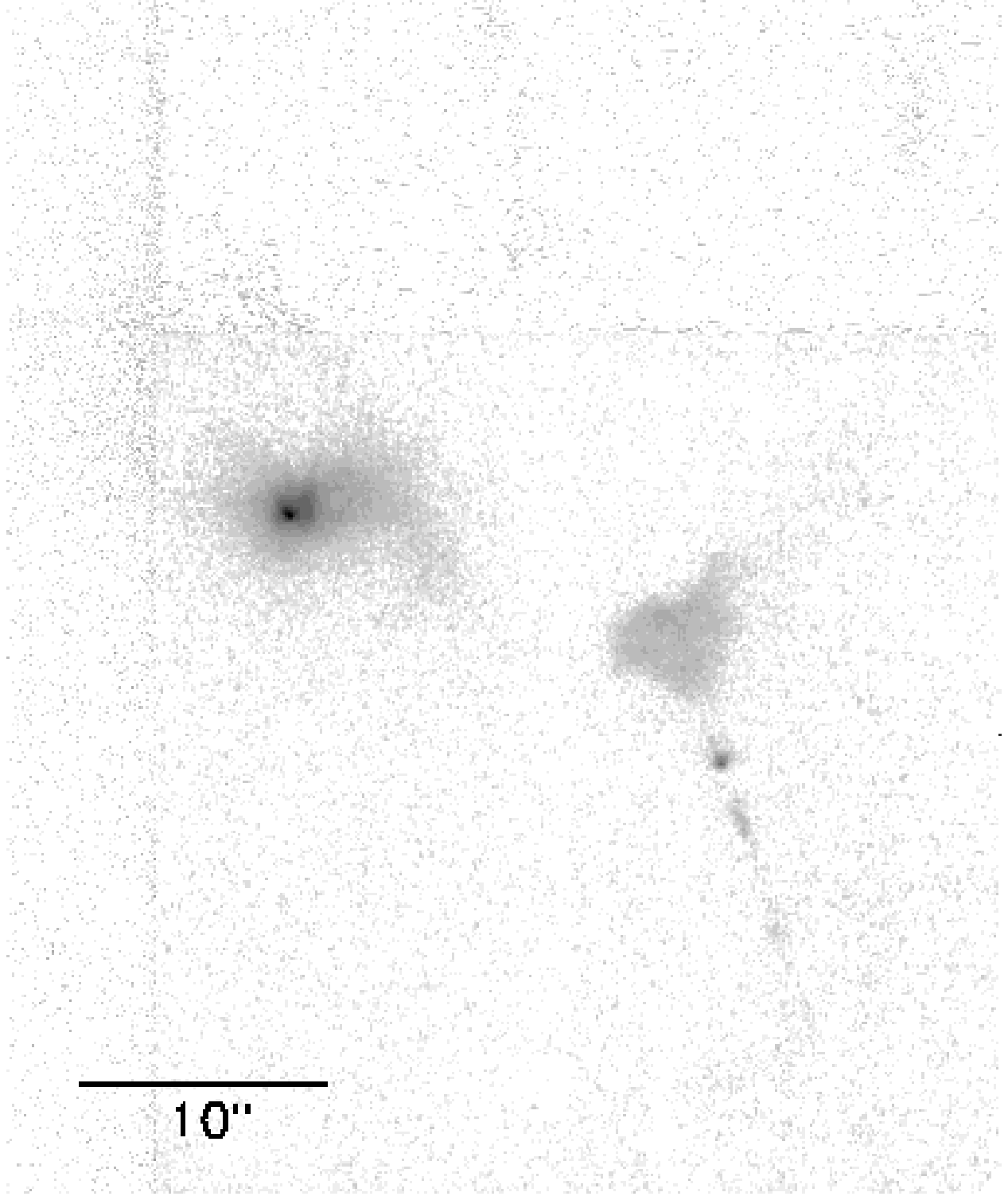}
\caption{\label{bilder} The FS Tau field in H$\alpha$ (left panel)
 and [S\,II] (right panel) without continuum subtraction. These images are
 mosaics of parts of the four quadrants of WFPC2, where the images obtained
 with the PC chip (lower right) have been re-scaled to fit the pixel scale of
 the three WF chips. The image size is $40\times 70\,\mathrm{arcsec}$.
 The nebulosity around FS~Tau~A and the structure denoted with R1 are
 reflection nebulae. The FS~Tau~B jet is in the upper part of the images,
 the faint counterjet is located southwest of FS~Tau~B.}

\end{figure}

\begin{figure}
\plotone{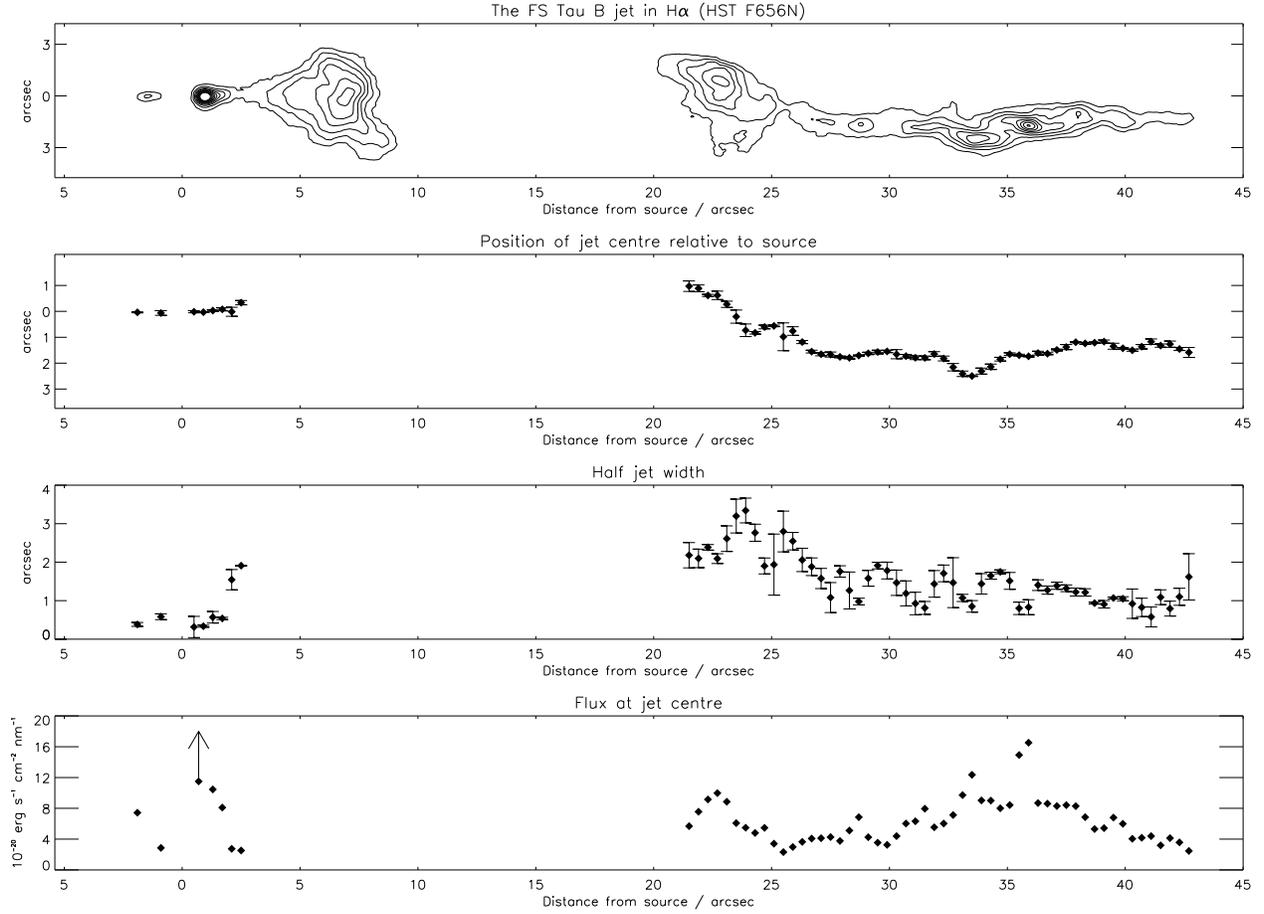}
\caption{\label{hamorph} The FS~Tau~B outflow in H$\alpha$. The
 first panel shows a contour plot of the jet (see Fig.\,\ref{bilder}).
 The second to fourth panels display the results of the Gaussian fits
 described in Sect.\,\ref{fstaub}: Position of jet centre, half jet width,
 and intensity at jet centre as function of distance to the source.}
\end{figure}

\begin{figure}
\plotone{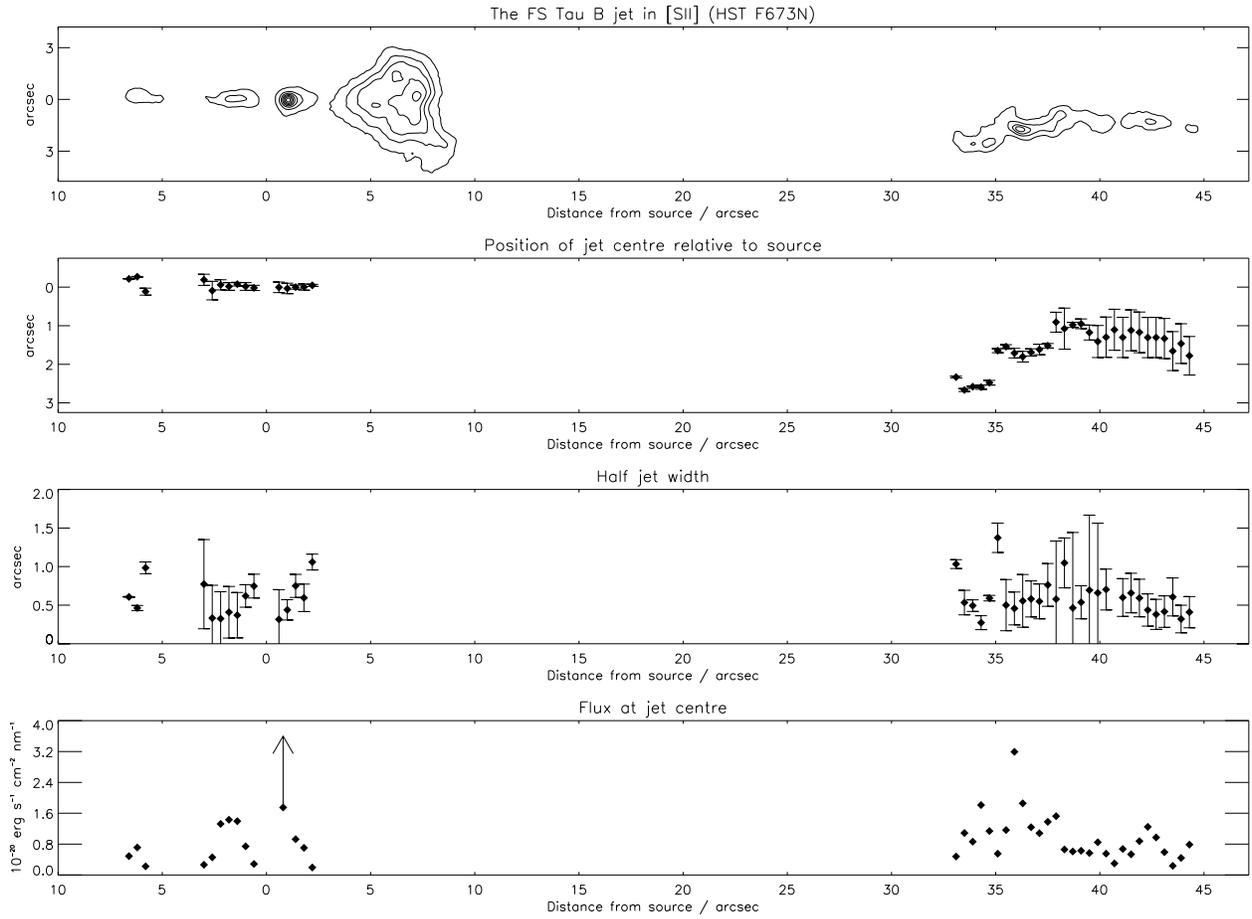}
\caption{\label{siimorph} The FS~Tau~B outflow in [S\,II].
 See caption of Fig.\,\ref{hamorph} for further explanation.}
\end{figure}

The intensity peaks, which occur in the jet after its
first reappearance (region A in Fig.\,\ref{bilder}), correspond to minima
of the jet width. Ray et al.\,(1996) have observed the HH 30 jet with HST
WFPC2 and found a similar behaviour for that outflow. So this may appear
as a common phenomenon if a sufficiently high spatial resolution is used.
The mentioned jet knots in the FS~Tau~B jet have circular or elliptical
shape, they do not exhibit the morphology of resolved bows. With respect
to the recollimation scenario discussed above they may represent regions
where the magnetic hoop stresses are maximum (O'Sullivan \& Ray 2000).
Another possible explanation is that this
fine structure is due to temporal and spatial variations within the jet
flow itself. In this case the flow would have a ``core'' region with
bright knots surrounded by another region with lower and roughly constant
flux density. Such a situation could be explained by the x-wind model of
young stellar objects (Shu et al. 2000 and references therein): This
theory predicts that only about one half of the jet flow is within a small
angle of some degrees around the polar axis while the rest exits in a wide
angle wind. A decision between these scenarios cannot be made solely on
the basis of our data. For example the proper motion measurements
of these small jet knots are required and this cannot be done by comparison
with previous ground-based imaging (e.\,g. EM98) because these images had a
too low spatial resolution to resolve morphological structures on comparable 
size scales. In addition high spatial resolution mm studies are required
to determine the characteristics and the kinematics of the ambient
environment surrounding FS Tau B. Such observations will have to await
the availability of ALMA.\\

Finally the jet appears much more collimated in [S\,II] than in H$\alpha$
where the jet width is about a factor two larger. This means
that the line ratio H$\alpha$/[S\,II] is increasing with lateral distance
from the jet centre. A similar situation has been found in HST WFPC2 images
of the HH~46/47 jet obtained by Heathcote et al.\,(1996). In this outflow the
body of the jet is primarily seen in [S\,II] emission while the H$\alpha$
emission merely denotes wisps and filaments at the border of the jet.
The finding that the higher excitation line occurs along the edges of the
jet suggests that a large fraction of the observed
emission is produced by entrainment of ambient material (Hartigan et
al 1993).

%% If you wish to include an acknowledgments section in your paper,
%% separate it off from the body of the text using the \acknowledgments
%% command.

%% Included in this acknowledgments section are examples of the
%% AASTeX hypertext markup commands. Use \url without the optional [HREF]
%% argument when you want to print the url directly in the text. Otherwise,
%% use either \url or \anchor, with the HREF as the first argument and the
%% text to be printed in the second.

\acknowledgments
J. E. and J. W. acknowledge support by the Deutsches Zentrum f\"ur Luft- und
Raumfahrt (grant number 50 OR 0009). We wish to thank Adam Frank and
Tom Gardiner for useful discussions on jet collimation mechanisms.
This paper is based on observations made with NASA/ESA {\it Hubble Space
Telescope}, obtained at the Space Telescope Science Institute, operated
under NASA contract No. NAS5-26555.

\clearpage

%% Use the figure environment and \plotone or \plottwo to include 
%% figures and captions in your electronic submission.

\clearpage

%% If you are not including electonic art with your submission, you may
%% mark up your captions using the \figcaption command. See the 
%% User Guide for details.
%%
%% No more than seven \figcaption commands are allowed per page, 
%% so if you have more than seven captions, insert a \clearpage 
%% after every seventh one. 

%% Tables should be submitted one per page, so put a \clearpage before
%% each one.

%% Two options are available to the author for producing tables:  the
%% deluxetable environment provided by the AASTeX package or the LaTeX
%% table environment.  Use of deluxetable is preferred.
%%

%% Three table samples follow, two marked up in the deluxetable environment,
%% one marked up as a LaTeX table.

%% In this first example, note that the \tabletypesize{}
%% command has been used to reduce the font size of the table.
%% Note also that the \label command needs to be placed 
%% inside the \tablecaption.

\clearpage

\end{document}